# On the Conservation of Physical Quantities

Douglas M. Snyder

ABSTRACT

Certain transformations of isolated physical systems underlying various conservation laws in physics are noted. As regards each of these transformations, there is a theoretical action that is *equivalent* to a matched physical action on the system. Evidence supporting this thesis is found in experimental psychological research where results of various experiments allow that the imagined rotation of an object is analogous to a physical rotation. The conservation laws based on the transformations noted are theoretical or mental in nature to the same extent that they are physical. When one of the conservation principles noted is tested by carrying out the relevant transformation on the system, this test is essentially one of the influence of mental activity on physical reality.

TEXT

A conservation law in physics, i.e., a law noting a zero rate of change over time for some physical quantity, reflects an underlying invariance of physical law under some transformation of an isolated physical system. Perhaps the most important conservation law is that concerning energy; this law reflects the invariance of an isolated physical system with regard to time. The conservation of linear momentum reflects the invariance for such a system with regard to spatial displacement. The invariance with regard to spatial rotation of an isolated physical system is the foundation for the conservation of angular momentum. Each of the spatial or temporal transformations underlying the conservation laws of energy, linear momentum, and angular momentum possesses a special nature given the relative character of physical events. There is no absolute origin for either space or time but only a chosen relative origin. Maintaining a set of axes while carrying out a physical action to transform the system is equivalent to transforming the axes while leaving the system untouched. For example, the act of moving a physical system along an axis $x$ representing a linear spatial direction from $x_1$ to $x_2$ ($x_2 > x_1$) is equivalent to leaving the system untouched and linearly moving the axis such that the coordinate of the system along this axis changes from $x_1$ to $x_2$. The movement of this axis is essentially a theoretical, or mental, action. Moving a measuring device representing the axis such that the system's coordinate changes from $x_1$

# On the Conservation

to $x_2$ is equivalent to adding $x_2 - x_1$ to all points on the axis and leaving the measuring device untouched. For each of the transformations noted, there is a theoretical action that is *equivalent* to a matched physical action on the system. The conservation laws based on these transformations are theoretical or mental in nature to the same extent that they are physical. The intrinsic relationship between mental activity and physical reality has also been discussed (2, 3).

Evidence supporting this thesis is found in experimental psychological research (1). They reported the results of various experiments, concluding that the imagined rotation of an object is analogous to a physical rotation. In one of Cooper's experiments, for example, subjects were asked to identify *whether a* figure in a particular spatially rotated orientation (without having witnessed the possible rotation of the figure) was the identical figure initially presented or a reflected (mirror image) version of this figure. The results indicated a linear dependence between the angle of rotation from the original position and the time subjects took to indicate that the figures were the same. For the reflected figures, a constant time factor was added to the time for identification of the figure; the linear dependence was preserved. This linear dependence is consistent with the time taken by subjects to engage in the proposed imagined rotation. Cooper and Shepard (1) maintained that imagined spatial rotations are internal representations developed as a result of their evolutionary adaptive significance in coping with the external physical world in which actual spatial rotations occur. But, as discussed, there is no basis for discriminating between the imagined rotation and the actual rotation in the formulation of the conservation of angular momentum. Mentally rotating a set of coordinate axes applied to a physical system is essentially an imagined physical rotation of the system. When one of the conservation principles noted is tested by carrying out the relevant transformation on the system, this test is essentially one of the influence of mental activity on physical reality.